\documentclass[sigconf]{acmart}

\usepackage{enumitem}
\usepackage{booktabs}  
\usepackage{caption}    
\usepackage{multirow}   
\usepackage{tabularx}
\usepackage{bm}
\usepackage{array}
\usepackage{appendix}
\usepackage{fvextra}
\usepackage{fancyvrb}
\usepackage{acmart-taps}

\DefineVerbatimEnvironment{PromptBlock}{Verbatim}{
  breaklines=true,
  breakanywhere=true,
  fontsize=\small,
  frame=single,
  xleftmargin=1em,
  xrightmargin=1em
}

\newcommand{\sys}[0]{\textsc{RemiAssist}}
\newcommand{\rev}[1]{\textcolor{black}{#1}}
\newcommand{\xsc}[1]{\textcolor{black}{#1}}
\newcommand{\sta}[1]{\textcolor{black}{#1}}

\AtBeginDocument{%
  }

\copyrightyear{2026}
\acmYear{2026}
\setcopyright{cc}
\setcctype{by-nc-nd}
\acmConference[UIST '26]{The 39th Annual ACM Symposium on User Interface Software and Technology}{November 02--05, 2026}{Detroit, MI, USA}
\acmBooktitle{The 39th Annual ACM Symposium on User Interface Software and Technology (UIST '26), November 02--05, 2026, Detroit, MI, USA}
\acmDOI{10.1145/3830398.3830485}
\acmISBN{979-8-4007-2856-3/2026/11}

\begin{document}

\title[\sys{}]{\sys{}: A Therapist-Supporting System for Photo-Based Reminiscence Therapy in Dementia Care}

\author{Shuchang Xu}
\authornote{This work was partially conducted during an academic visit to the MIT Media Lab.}

\affiliation{
\institution{The Hong Kong University of Science and Technology}
\city{Hong Kong}
\country{China}
}
\orcid{0000-0002-7642-9044}
\email{sxuby@connect.ust.hk}

\author{Minglong Tang}
\affiliation{
\institution{East China Normal University}
\city{Shanghai}
\country{China}
}
\orcid{0009-0009-5511-827X}
\email{minglong.hci@gmail.com}

\author{Junyan Mao}
\affiliation{
\institution{Fudan University}
\city{Shanghai}
\country{China}
}
\orcid{0009-0009-5511-827X}
\email{jymao25@m.fudan.edu.cn}

\author{Xiaofu Jin}
\authornote{Corresponding authors.}
\affiliation{
\institution{Univeristy of Stuttgart}
\city{Stuttgart}
\country{Germany}
}
\orcid{0000-0002-7239-3769}
\email{xiaofu.jin@vis.uni-stuttgart.de}

\author{Wazeer Zulfikar}
\affiliation{%
  \institution{MIT Media Lab}
  \city{Cambridge, MA}
  \country{USA}}
\email{wazeer@media.mit.edu}

\author{Yasith Samaradivakara}
\affiliation{%
  \institution{MIT Media Lab}
  \city{Cambridge, MA}
  \country{USA}}
\email{yasith@media.mit.edu}

\author{Jiayi Zhou}
\affiliation{%
\institution{The Hong Kong University of Science and Technology}
\city{Hong Kong}
\country{China}}
\email{jzhoudp@connect.ust.hk}

\author{Huamin Qu}
\affiliation{
\institution{The Hong Kong University of Science and Technology}
\city{Hong Kong}
\country{China}
}
\orcid{0000-0002-3344-9694}
\email{huamin@cse.ust.hk}

\author{Yuling Sun}
\authornotemark[2]
\authornote{Silver-X MOE Philosophy \& Social Sciences Laboratory, Fudan Institute on Aging.}
\affiliation{
\institution{Fudan University}
\city{Shanghai}
\country{China}
}
\orcid{0000-0003-1726-5913}
\email{yulingsun@fudan.edu.cn}

\author{Pattie Maes}
\affiliation{
\department{}
\institution{MIT Media Lab}
\city{Cambridge, MA}
\country{USA}
}
\orcid{0000-0002-7722-6038}
\email{pattie@media.mit.edu}

\renewcommand{\shortauthors}{Xu et al.}

\sloppy

\begin{abstract}

Despite growing interest in applying AI to photo-based reminiscence therapy (PRT) for people with dementia (PwD), 
existing systems primarily focus on PwD-AI interaction and often overlook therapists' critical role in practical PRT delivery. 
\rev{We present \sys{}, a system that supports therapist-in-the-loop PRT through AI-assisted planning and real-time facilitation.} 
\sys{} incorporates two core techniques: 
(1) a \textit{Memory Graph}, which organizes key life events from a PwD's photo collection into a hierarchical graph to support theme-centered intervention planning; and 
(2) a \textit{Context-Aware Guiding Strategy}, which provides real-time suggestions to help therapists guide reminiscence conversations and respond to sensitive situations. 
A field study with eight therapist–PwD dyads \sta{suggests that \sys{} was associated with a 44\% improvement in planning efficiency, a 54\% increase in conversation duration, and timely support for handling sensitive situations.} 
We highlight opportunities for AI systems to empower therapists and enable more personalized reminiscence therapy in dementia care.

\end{abstract}



\begin{teaserfigure}
\centering
  \includegraphics[width=1.0\textwidth]{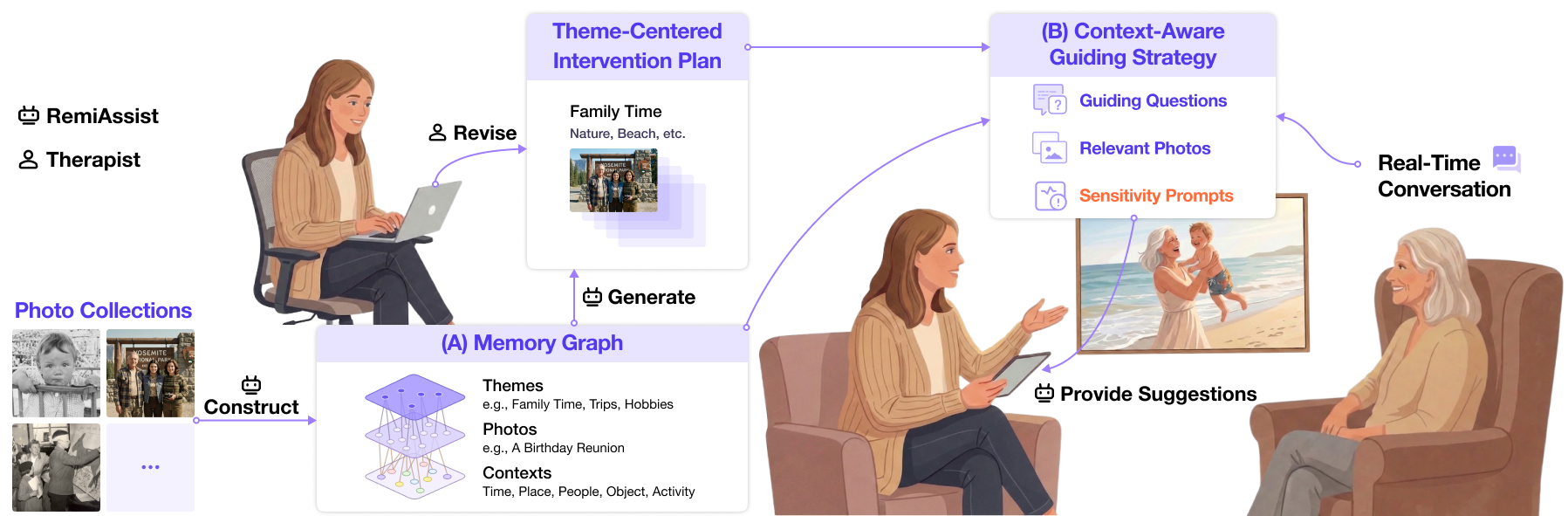}
  \caption{
    \rev{\sys{} supports \textit{therapist-in-the-loop} photo-based reminiscence therapy in dementia care.} 
    It incorporates two core techniques: 
    (A) a \textit{Memory Graph}, which organizes key life events and their contextual relationships in a hierarchical graph; and 
    (B) a \textit{Context-Aware Guiding Strategy}, which provides real-time suggestions to help therapists facilitate reminiscence conversations and respond to sensitive situations.
  }
  \Description{}
  \label{fig:teaser}
\end{teaserfigure}

\maketitle


\section{Introduction}

Dementia significantly impacts the quality of life of older adults and places substantial emotional and caregiving burdens on their families and therapists \cite{world2021global,who_dementia_2025}. 
Reminiscence therapy (RT), which typically uses memory cues such as old objects and personal photos to guide people with dementia (PwD) in recalling and sharing past life experiences~\cite{woods2018reminiscence}, is widely recognized as an effective psychosocial intervention for PwD, with demonstrated benefits for mood, social engagement, and psychological well-being \cite{woods2018reminiscence,subramaniam2012impactRT,huang2015reminiscence}.

In practice, delivering RT effectively requires substantial effort and expertise from therapists \cite{NPI_chi24,lazar2014systematic,messmer2025designing}. Specifically, RT is inherently personalized and situated, requiring therapists not only to prepare appropriate intervention materials in advance, but also to skillfully facilitate reminiscence conversations in real time during sessions \cite{woods2018reminiscence,cuevas2020reminiscence}. Moreover, RT for PwD is ethically sensitive, requiring therapists to recognize emotionally vulnerable moments and provide timely support when necessary \cite{bittersweet_chi25}. As a result, therapists play a pivotal role in ensuring the effective, ethical, and emotional delivery of RT \cite{NPI_chi24}. However, a severe shortage of well-trained therapists in real-world settings limits the \xsc{accessibility} of RT for the broader PwD population \cite{kuwahara2006networked,sun2023care}.

Prior research has explored the use of AI to support RT for PwD \cite{morales2021conversational,hong2024conect}. However, most existing systems (e.g. \cite{mindtalker_chi24,morales2021conversational,auto_RT_icmr20,sun2025chorus}) primarily focus on direct interaction between PwD and AI, overlooking the central role of therapists in practical RT delivery \cite{seah2026rememo}. More recent work instead advocates for a shift in design orientation --- from replacing therapists to supporting their workflows and professional practices \cite{seah2026rememo,NPI_chi24}.

Building on this perspective, we present \sys{}, \rev{a system that supports therapist-in-the-loop PRT through AI-assisted planning and real-time facilitation.} 
\sys{} focuses on photo-based RT (PRT), as photos are one of the most commonly used memory cues in practice \cite{wu2020role,jiang2021library}. 
To inform our design, we first conducted a formative study with therapists and identified three key needs: 
preparing theme-centered intervention plans, 
guiding real-time conversations, and 
monitoring sensitive conditions to ensure safe and ethical delivery. 
To address these needs, \sys{} incorporates two core techniques. 
First, a {\textit{Memory Graph}} organizes key life events from a PwD's photo collection into a hierarchical graph, enabling therapists to prepare theme-centered intervention plans more effectively. 
Second, a {\textit{Context-Aware Guiding Strategy}} supports real-time session facilitation by providing therapists with contextually relevant conversational prompts and photo suggestions based on PwD's evolving interests. 
It also identifies emotionally sensitive situations and suggests alternative topics to help therapists keep the reminiscence process both safe and engaging.

To evaluate \sys{}, we conducted a within-subject field study with eight therapist-PwD dyads, comparing \sys{} with a baseline system reflecting therapists' manual PRT practice. 
Therapists used both systems to plan and facilitate two comparable PRT sessions. 
\sta{The results suggest that \sys{} was associated with a 44\% improvement in planning efficiency, a 54\% increase in conversation duration, and lower cognitive load among therapists during facilitation.} 
Therapists further reported that the context-aware suggestions were particularly valuable for redirecting conversations during distressing moments. 
Based on these findings, we highlight opportunities for context-aware systems to augment therapist-PwD communication and enable safer and more personalized delivery of reminiscence therapy in dementia care.

In summary, our contributions are threefold:
\begin{itemize}[leftmargin=*, labelindent=0pt, itemindent=0pt]
    \item We present \sys{}, a system that supports \textit{therapist-in-the-loop} reminiscence therapy in dementia care. It incorporates a \textit{Memory Graph} for theme-centered intervention planning and a \textit{Context-Aware Guiding Strategy} for real-time facilitation.
    \item \sta{We conduct a field study that provides preliminary evidence} of how \sys{} supports therapists in planning and facilitating real-world RT sessions with PwD ($N = 8$ dyads).
    \item We derive design implications for context-aware systems that augment therapist-PwD communication and enable more personalized reminiscence therapy in dementia care.
\end{itemize}



\section{Related Work}
Our work builds on three core research areas: 
(1) technologies for reminiscence therapy, 
(2) personal memory augmentation, and 
(3) context-aware systems in conversations.

\subsection{Technologies for Reminiscence Therapy}

Reminiscence therapy (RT) is a widely used cognitive and psychosocial intervention for people with dementia (PwD) \cite{woods2018reminiscence}. In practice, RT typically relies on memory cues, such as photos, videos, music, and old objects, to evoke memories of past experiences \cite{woods2018reminiscence,sun2025chorus}, supporting PwD in recalling, sharing, and reflecting on these experiences through guided conversation. Prior research on technology-supported RT has developed along two main directions \cite{seah2026rememo}: (1) tools for preparing memory-elicitation materials and (2) AI systems for independent reminiscence. 
The first direction focuses on providing multimedia stimuli, such as photos \cite{residential_care_chi24,wang2024promoting}, videos \cite{kuwahara2006networked}, music \cite{music_RT_chi24}, and virtual reality experiences \cite{remiverse_imwut25,relive_vr2025}, to elicit memory recall. 
For example, \textit{RemiHaven} uses AI-generated images to recreate past scenes from older adults' narration \cite{remihaven_chi25}, 
while \textit{RemVerse} employs virtual reality to immerse older adults in nostalgic environments \cite{remiverse_imwut25}. 
The second line of research focuses on designing conversational AI \cite{sun2025chorus,mindtalker_chi24} or social robots \cite{bittersweet_chi25,cruz2020social} to guide and facilitate RT. For instance, \textit{RemiBuddy} uses a multi-agent conversational system to guide older adults in sharing past memories \cite{sun2025chorus}, and \textit{MindTalker} uses a conversational agent to support people with early-stage dementia in discussing personal experiences \cite{mindtalker_chi24}. 

While these systems have advanced the development of technology-mediated RT, they have been shown to face significant usability and effectiveness challenges in real-world settings \cite{NPI_chi24,bittersweet_chi25}. In particular, they risk oversimplifying the therapeutic process and overlooking the emotional and social dynamics during reminiscence \cite{seah2026rememo,de2024co}. 
Recent work therefore argues for a shift in design orientation: rather than replacing therapists, AI should be developed to support their workflows and professional practice \cite{NPI_chi24,relationship_dynamics_chi25,seah2026rememo}. 
\xsc{
Yet, how AI can provide meaningful assistance while preserving therapists' professional judgment remains an open challenge \cite{NPI_chi24,relationship_dynamics_chi25}. 
Our work addresses this challenge across two stages of RT. Before sessions, the system generates editable plans that therapists can refine based on their clinical expertise. 
During sessions, a glanceable interface surfaces context-aware suggestions that therapists can selectively incorporate, thereby preserving their control over therapeutic decisions.
}

\subsection{Personal Memory Augmentation}
A large body of HCI research has explored personal memory augmentation systems, including technologies for capturing daily experiences \cite{hodges2006sensecam}, retrieving past events \cite{li2025omniquery,mempal_iui25,memoro_chi24}, and supporting reminiscence \cite{memory_reviver,peesapati2010pensieve}. 
Early work such as \textit{SenseCam} used wearable cameras to record daily life and support later recall \cite{hodges2006sensecam}. 
More recent systems have expanded beyond visual capture alone. 
For example, \textit{Memoro} continuously records conversations through headphones to enable on-demand memory retrieval \cite{memoro_chi24}, 
while \textit{MemPal} uses wearable cameras to capture older adults' daily activities and help them recall objects in the home \cite{mempal_iui25}. 
\textit{OmniQuery} further applies retrieval-augmented generation to personal photo collections, allowing users to query past experiences more effectively \cite{li2025omniquery}. 

Despite these advances, existing systems primarily focus on supporting \textit{retrieval}: helping users recover specific memories after they have been recorded \cite{li2025omniquery}. 
When applied to practical photo-based RT, however, they fail to support another critical stage: structured \textit{planning}, in which therapists need to identify meaningful life themes, select relevant photos, and arrange them into a coherent sequence for discussion \cite{jiang2021library,RT_handbook_2005}. 
This process requires understanding the PwD's experiences at multiple levels of detail while also recognizing connections across events. 
To address this need, we designed a \textit{Memory Graph} that is automatically constructed to represent PwD's life experiences and their relationships, thereby helping therapists organize photo archives into coherent session plans.

\subsection{Context-Aware Systems in Conversations}
The HCI community has long explored context-aware AI systems that support human-human conversations (e.g., \cite{visualcaption,crosstalk,search_agent_dis18,smartwatch_iswc20}). 
One major line of work augments everyday conversations by surfacing contextually relevant information. 
For example, Andolina et al. developed a proactive search agent that monitors ongoing dialogue, identifies mentioned keywords, and presents relevant web results in real time \cite{search_agent_dis18}. Similarly, Ogawa et al. introduced a smartwatch-based system that suggests conversation topics to facilitate social interaction \cite{smartwatch_iswc20}. 
A second line of research has examined context-aware AI in mental health communication. 
For instance, in email counseling, \textit{CAIA} helps therapists review long email threads by generating case summaries \cite{fu2023enhancing}. 
Similarly, in online mental health peer-support platforms, \textit{HAILEY} provides AI-generated feedback to help peer supporters compose more empathic responses \cite{sharma2023human}. 
These systems demonstrate the value of context-aware assistance in emotionally sensitive conversations. 
However, they primarily support online, text-based communication, where users have time to reflect, revise, and respond asynchronously. 
By contrast, face-to-face therapeutic conversations impose distinct real-time demands. 
Therapists must interpret evolving context and make decisions in real time \cite{martin2024counseling}. 
These demands are especially pronounced in therapist-PwD communication during RT. Because of cognitive decline, PwD may struggle to sustain attention, drift off topic, or become distressed during conversation \cite{kindell2017everyday}. 
At the same time, therapists must continuously adapt their guidance to the PwD's interests, responses, and emotional state, which requires both facilitation expertise and substantial cognitive effort \cite{NPI_chi24,de2024co}. 
\sys{} addresses these challenges by providing context-aware suggestions that therapists can quickly review during sessions. 
It further detects sensitive moments to support safe and timely intervention. 



\section{Formative Study}

We conducted a formative study with five therapists (F1-F5 in Table~\ref{tab:formative_demographics}) to identify the key challenges and design opportunities in real-world photo-based reminiscence therapy. 
Participants were recruited from dementia care institutions specializing in dementia intervention, \xsc{with criteria of having at least one year of experience in regularly conducting PRT in dementia care.} 
We conducted semi-structured interviews with each therapist to examine three key aspects: 
(1) their current practices in conducting PRT, 
(2) key challenges encountered during PRT sessions, and 
(3) expectations for AI support within their workflow. 
Therapists were encouraged to recall prior sessions and share materials they had previously used to illustrate their practices and challenges. 
To provide a concrete reference for discussing AI support, we introduced examples of conversational AI tools such as ChatGPT\footnote{https://chatgpt.com}. 
All interviews were conducted in person, with each session lasting about 40 minutes. 
We analyzed the interview data using thematic analysis \cite{clarke2017thematic}. 
\xsc{Two authors independently coded transcribed data, compared initial codes, and resolved discrepancies through iterative discussion. They then grouped these codes into higher-level themes and reviewed these themes to ensure that they accurately captured the key findings.} 
From this analysis, we derived four key design considerations for supporting therapists in conducting PRT with PwD:

\textbf{(DC1) Organize intervention plans around meaningful life themes.} 
Effective PRT sessions are typically grounded in life themes (e.g., family and hobbies) that are personally meaningful to each PwD, with photos and prompts selected accordingly. 
However, preparing these personalized materials is often labor-intensive. Systems should therefore help therapists in organizing intervention materials, generating personalized intervention plans, and reducing the preparation workload.

\textbf{(DC2) Provide contextual knowledge to support photo interpretation.} 
Therapists need a rich understanding of the PwD's personal background to effectively facilitate PRT. In practice, however, obtaining such information is often time-consuming and effortful, and in many cases impractical because family members may be unavailable to recall relevant information. 
Systems should therefore provide contextual cues, such as likely time, location, and landmarks, to help therapists better understand the PwD’s life history and interpret photos used in therapy.

\textbf{(DC3) Support adaptive guidance based on real-time interests.} 
Although therapists may prepare sessions in advance, reminiscence conversations are highly dynamic and situated, often unfolding beyond the planned materials, which creates challenges for therapists in sustaining coherent and responsive guidance in real time. 
Systems should therefore support therapists in adapting their guidance to the PwD's real-time interests. This includes suggesting follow-up prompts or surfacing additional photos that align with topics emerging during the conversation.

\textbf{(DC4) Monitor sensitive conditions to ensure safe and ethical intervention.} 
Due to cognitive decline, PwD may have difficulty sustaining attention and drift off topic \cite{kindell2017everyday}. 
Reminiscence may also trigger distressing memories, including divorce, bereavement, or other major life disruptions \cite{bittersweet_chi25}, which may introduce emotional and ethical risks into the therapeutic process. 
Systems should therefore help therapists recognize these sensitive conditions so they can respond appropriately and ensure a safe therapeutic process.



\section{\sys{}}

\begin{figure*}[!htbp]
    \centering
    \includegraphics[width=1.0\linewidth]{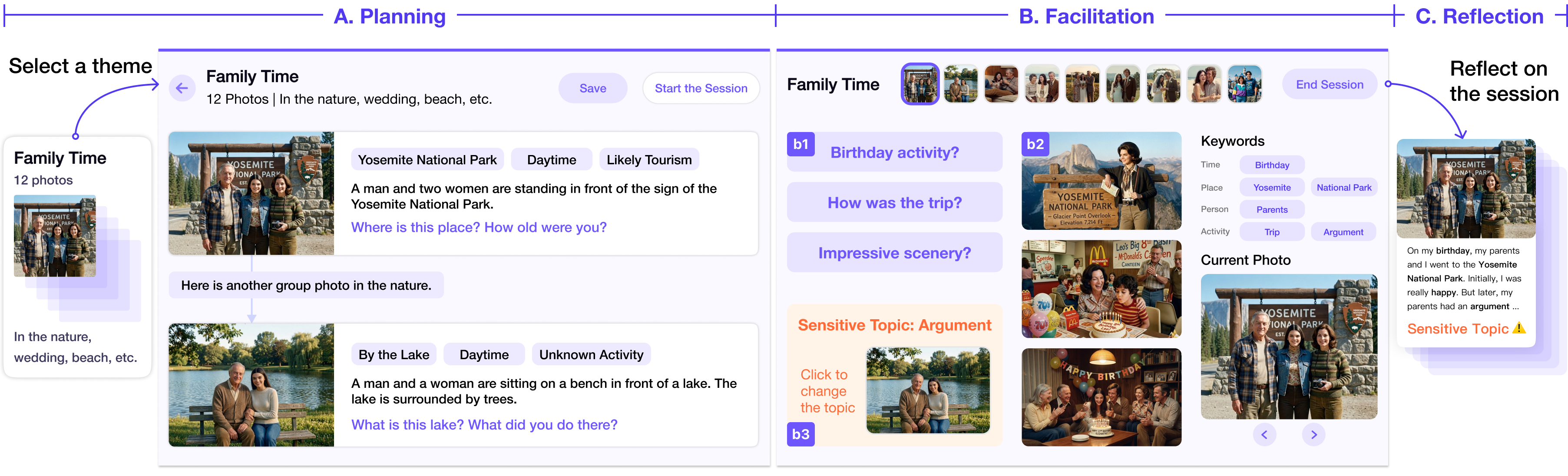}
    \caption{
    \sys{} supports therapists in \textit{planning}, \textit{facilitating}, and \textit{reflecting} on each PRT session. 
    (A) During planning, \sys{} generates theme-centered session plans, which therapists can further revise based on their professional judgment. 
    (B) During facilitation, \sys{} provides therapists with three types of context-aware suggestions: 
    (b1) \textit{guiding questions} to sustain the conversation, 
    (b2) \textit{relevant photos} to adapt to the PwD's real-time interests, and 
    (b3) \textit{sensitivity prompts} to respond to sensitive situations. 
    These suggestions are designed for quick, at-a-glance review, minimizing intrusion to the natural flow of the conversation. 
    (C) During post-session reflection, \sys{} summarizes key life events and associated emotional responses to support future intervention planning.
    }
    \label{fig:walkthrough}
\end{figure*}

Based on the identified DCs, \rev{we present \sys{}, a system that supports therapist-in-the-loop PRT in dementia care.} 
Specifically, \sys{} supports therapists in planning and facilitating PRT sessions through two core techniques: 
(1) a \textbf{\textit{Memory Graph}}, which organizes key life events from a PwD's photo collection into a hierarchical graph to support efficient intervention planning (addressing \textbf{DC1}--\textbf{DC2}), and 
(2) a \textbf{\textit{Context-Aware Guiding Strategy}}, which provides therapists with real-time, adaptive suggestions during session facilitation  (addressing \textbf{DC3}--\textbf{DC4}). 
In the following, we first present a user walkthrough (Section~\ref{sec:walkthrough}) and then describe these two techniques in detail (Sections~\ref{sec:memory_graph}--\ref{sec:guiding_strategy}).

\subsection{User Scenario Walkthrough}\label{sec:walkthrough}

To illustrate how \sys{} works in practice, we follow Eva, a therapist, as she conducts PRT with Lisa, a person with dementia. 
\sys{} supports Eva throughout the PRT workflow by helping her \textit{plan}, \textit{facilitate}, and \textit{reflect} on each session.

\textbf{Pre-Session Planning.} 
Before the session, Eva uses \sys{} to prepare a session plan. 
She starts by browsing a list of \textit{reminiscence themes} provided by \sys{}, such as childhood, family, and hobbies, and selects \emph{family time} as the focus of the session. 
\sys{} then provides a plan for the selected theme 
(see Figure~\ref{fig:walkthrough}A), including a list of relevant photos. 
Each photo is accompanied by inferred contextual details, such as the likely time and place, helping Eva understand its background. 
The system also suggests guiding questions to help Eva initiate conversations. 
After reviewing the plan, Eva adds several questions according to her professional experience and finalizes the plan for the session.

\textbf{Real-Time Facilitation.} 
During the session, Eva delivers the intervention based on the plan. 
She begins by showing Lisa the first photo and asking the question, ``\textit{Where is this place?}'' 
Lisa responds, ``\textit{This was at the Yosemite National Park. I was very happy because it was my birthday ...}'' 
As Lisa speaks, \sys{} analyzes the ongoing conversation and generates relevant follow-up questions 
(see Figure~\ref{fig:walkthrough}B). 
Eva notices that the prompt ``\textit{Birthday activity?}'' aligns naturally with Lisa's interest and uses it to deepen the discussion by asking, ``\textit{What did you do on your birthday?}'' 
Because these prompts are suggestive rather than prescriptive, Eva can ignore those that do not fit the flow of the conversation, thereby preserving her agency and professional judgment in guiding the session.

As Lisa recalls past events, she mentions a major argument with her parents. 
Recognizing that this topic may cause distress, Eva decides to gently redirect the conversation. 
She selects an alternative prompt suggested by \sys{}: ``\textit{Change the topic to [another photo]}''. 
These prompts are designed to be relevant to the session's theme while avoiding potentially distressing memories, 
thereby helping Eva shift the discussion toward a more comfortable topic.

\textbf{Post-Session Reflection.} 
After the session, \sys{} generates a summary of the memories shared by Lisa, along with her emotional responses (see Figure~\ref{fig:walkthrough}C). 
Eva reviews this summary, documents key life events, and identifies topics that Lisa is more comfortable discussing. 
These insights help Eva better understand Lisa's life histories and prepare for future sessions.

\subsection{Pipeline Overview}\label{sec:pipeline_overview}

\sys{} supports therapists in planning and facilitating PRT sessions through two core techniques: 
(1) a \textbf{\textit{Memory Graph}} for theme-centered intervention planning, and (2) a \textbf{\textit{Context-Aware Guiding Strategy}} for real-time facilitation. 
The following sections provide details for each technique.

\subsection{Memory Graph}\label{sec:memory_graph}

\xsc{Prior photo-memory systems typically organize memories along a linear timeline \cite{memory_reviver,li2025omniquery}. 
Yet, human autobiographical memories are often connected through shared contexts, including people, places, and activities \cite{lee2007providing,tulving1984precis}. 
To represent these connections, the \textit{Memory Graph} models key life events and their relationships as a non-linear graph. 
Its pipeline identifies associations at multiple levels of detail, supporting both theme-centered planning (\textbf{DC1}) and contextualized interpretation of a PwD's life history (\textbf{DC2}).
}


\subsubsection{\textbf{Memory Graph Structure}} 
\begin{figure}[!t]
    \centering
    \includegraphics[width=0.95\linewidth]{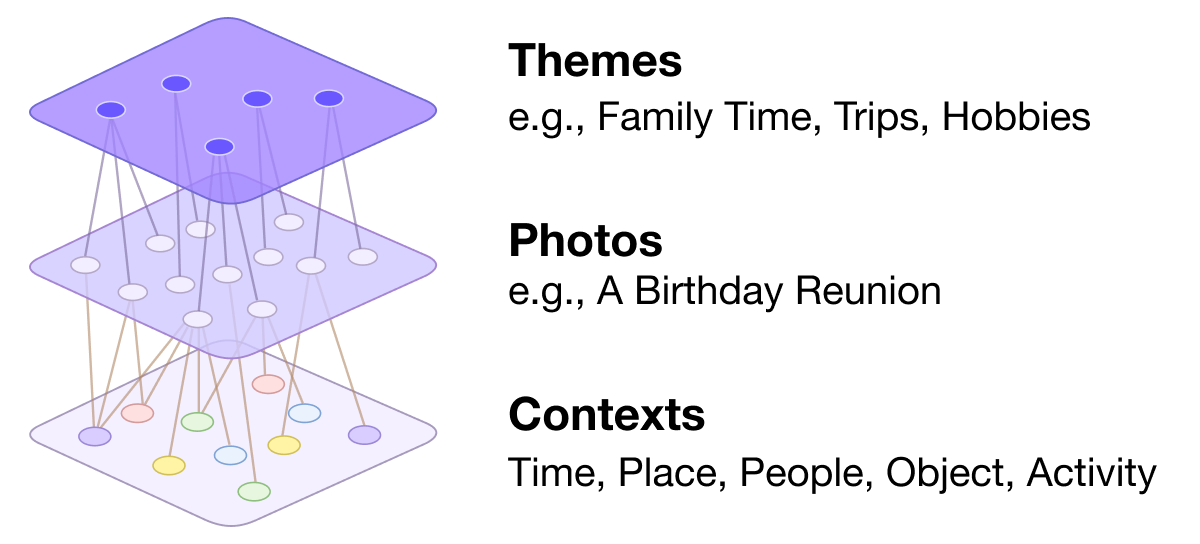}
    \caption{The structure of the \textit{Memory Graph}.}
    \label{fig:memory_graph}
\end{figure}

Figure~\ref{fig:memory_graph} shows the structure of the \textit{Memory Graph}, which comprises three hierarchical levels: 
``\emph{Themes} $\rightarrow$ \emph{Photos} $\rightarrow$ \emph{Contextual Details}''. This design is inspired by the hierarchical organization of human autobiographical memory \cite{conway1987organization,conway2000construction}. 
At the first level, \textbf{theme nodes} represent high-level themes of life experiences \cite{conway2005memory}, such as ``\textit{Family Time}''. 
At the second level, \textbf{photo nodes} represent specific experiences within each theme, such as ``\textit{a Birthday Reunion}''. 
At the third level, \textbf{context nodes} represent contextual details associated with each photo. 
Based on empirical studies of episodic memory \cite{lee2007providing,tulving1984precis,li2025omniquery}, contextual details are organized into five types: \emph{time}, \emph{place}, \emph{people}, \emph{object}, and \emph{activity}. 
Context nodes can be shared across multiple photos, thereby linking events that share similar contextual information.

\subsubsection{\textbf{Memory Graph Construction}} 
\begin{figure}[!t]
    \centering
    \includegraphics[width=1.0\linewidth]{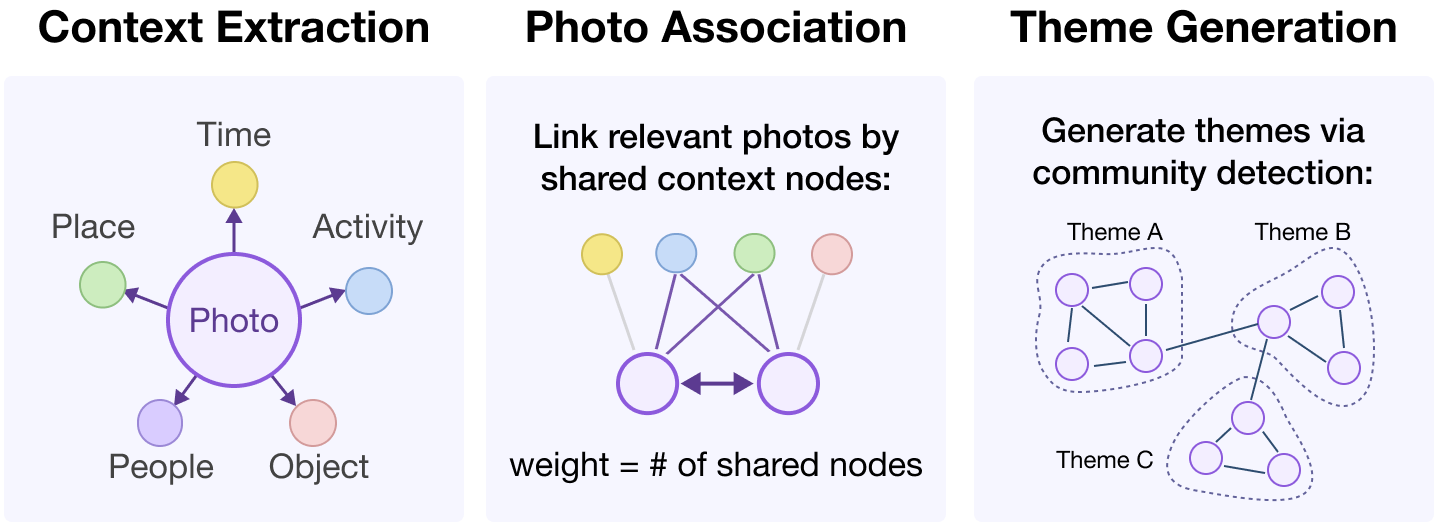}
    \caption{The pipeline for constructing the \textit{Memory Graph}.}
    \label{fig:memory_graph_construction}
\end{figure}

To operationalize this hierarchical representation from personal photo collections, we instantiate each level through a multi-stage construction pipeline.
Figure~\ref{fig:memory_graph_construction} illustrates the pipeline for constructing the \textit{Memory Graph}. 
Given a photo collection curated for reminiscence use, in which each photo typically corresponds to a salient life event, the pipeline builds the \textit{Memory Graph} in three stages: 
(1) context extraction, 
(2) photo association, and 
(3) theme generation.

\textbf{Context Extraction.} 
In the first stage, \sys{} extracts five types of contextual information from each photo: \textit{time}, \textit{place}, \textit{people}, \textit{objects}, and \textit{activities}. 
People are identified using InsightFace. 
The other four context types are inferred using Qwen-VL-Plus \cite{wang2024qwen2vl} following the OmniQuery method \cite{li2025omniquery}, which generates textual descriptions for each aspect. 
After context extraction, the system merges context nodes across photos to form a graph structure. Person nodes are merged based on \textit{visual identity}, with DBSCAN \cite{dbscan} used to cluster faces of the same individual. 
The remaining context nodes are merged based on \textit{semantic similarity}, using Qwen3-Max \cite{yang2025qwen3} for topic modeling and semantic grouping. 
This hybrid approach allows the system to capture both perceptual identity (for people) and semantic relationships (for other contextual elements).

\textbf{Photo Association.} 
In the second stage, \sys{} establishes links between photos based on contextual overlap. Specifically, two photos are connected if they share at least one context node, and the edge weight between them is defined as the number of shared context nodes. In this way, photos with greater contextual overlap are connected by stronger edges, enabling the graph to represent different degrees of event relatedness.

\textbf{Theme Generation.} 
In the final stage, \sys{} derives high-level reminiscence themes by clustering photos with strong contextual relevance. Specifically, it applies the Leiden community detection algorithm \cite{Leidentraag2019louvain} to the graph, partitioning it into densely connected communities with sparse inter-community links. Each community is treated as a reminiscence theme. Based on the shared contexts within each community, \sys{} uses Qwen3-Max \cite{yang2025qwen3} to generate a theme title, such as ``\textit{Family Time}.''
This separation allows structural grouping to be driven by graph connectivity while leveraging language models for interpretable theme labeling.

\subsubsection{\textbf{Intervention Plan Generation}}
After constructing the \textit{Memory Graph}, \sys{} maps each theme to a session plan automatically. 
Figure~\ref{fig:walkthrough}(A) illustrates an example. 
Photos within the plan are ordered so that adjacent images share similar contextual elements, which helps maintain a coherent conversation flow. For each photo, the system generates guiding questions, covering person, time, location, and activity in accordance with RT guidelines \cite{RT_handbook_2005}. 
The ordering and guiding questions are generated using a vision-language model (Qwen3-VL-Plus~\cite{wang2024qwen2vl}) with prompts that preserve contextual consistency across neighboring photos. 
The generated plan is intended as a draft for therapist review rather than a finalized version. 
Therapists can further refine the plan by editing any field, such as adding or removing photos, reordering them, revising guiding questions, or updating contextual details.

\subsection{Context-Aware Guiding Strategy}\label{sec:guiding_strategy}
During real-time facilitation, \sys{} provides therapists with context-aware suggestions to support adaptive guidance and ensure safe intervention (addressing \textbf{DC3}--\textbf{DC4}). 

\subsubsection{\textbf{Guiding Strategy Overview}} 
\begin{figure}[!t]
    \centering
    \includegraphics[width=1.0\linewidth]{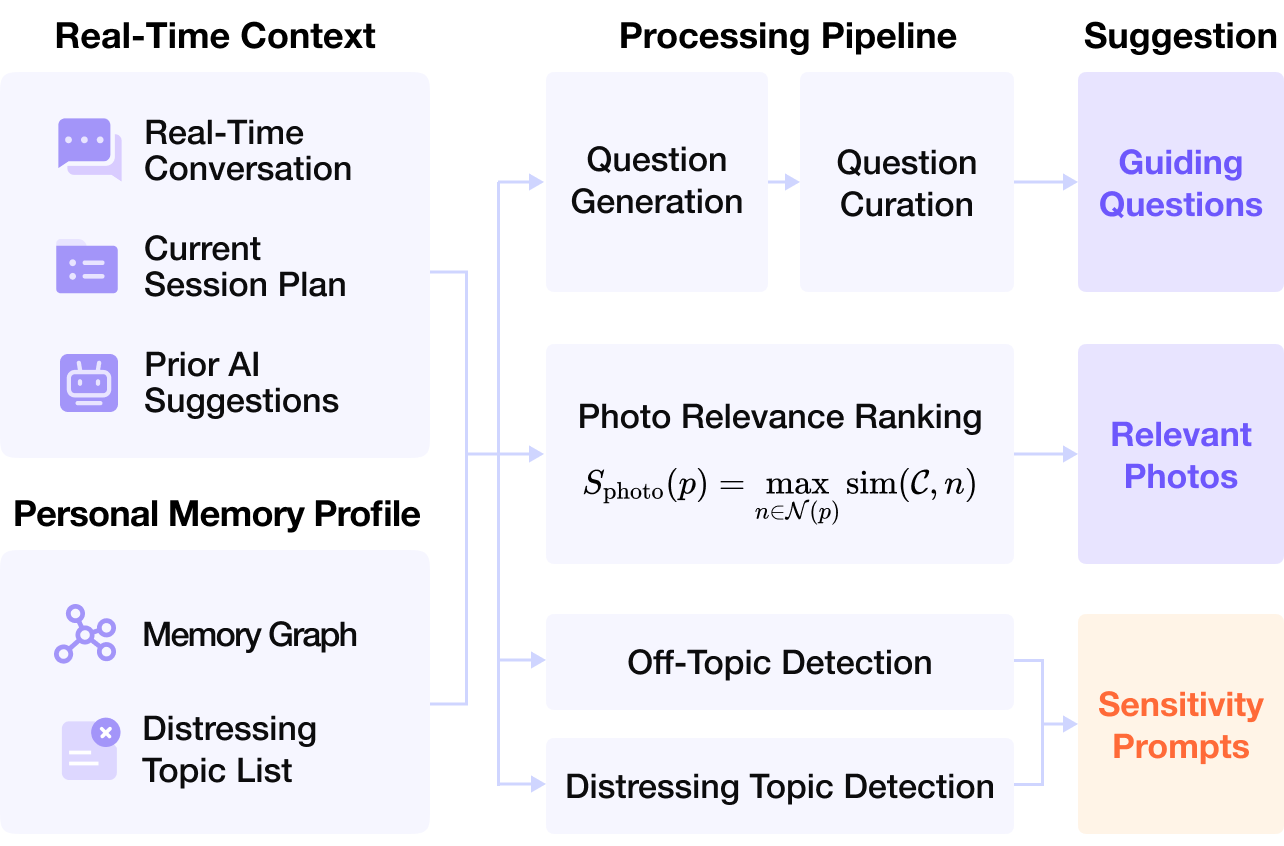}
    \caption{\sys{} generates three types of context-aware suggestions based on the real-time context and each PwD's memory profile.}
    \label{fig:guiding_strategy}
\end{figure}

\sys{} provides three types of context-aware suggestions during reminiscence sessions: 
(1) \textit{guiding questions} to sustain the conversation, 
(2) \textit{relevant photos} to adapt to the PwD's real-time interests, and 
(3) \textit{sensitivity prompts} to respond to sensitive situations. 
To ensure that these suggestions are \textit{adaptive} to the ongoing conversation and \textit{personalized} to each PwD's life history, \sys{} generates them based on both the real-time conversation context and the PwD's memory profile, as shown in Figure~\ref{fig:guiding_strategy}. 
The memory profile includes both the \textit{Memory Graph} and a list of distressing topics provided by family members, which helps the system avoid potentially sensitive topics.

Figure~\ref{fig:walkthrough}(B) shows the user interface for real-time facilitation. 
The interface is designed for \textit{at-a-glance} use, enabling therapists to review suggestions quickly without interrupting the flow of conversation. 
To support this lightweight interaction, the system presents suggestions in a continuously updated, auto-scrolling list that adapts to the ongoing dialogue. 
In the following sections, we describe each suggestion type in detail.

\subsubsection{\textbf{Guiding Questions}} 
Guiding questions are brief prompts that help therapists sustain the conversation. 
They are generated based on three principles drawn from RT guidelines \cite{RT_handbook_2005}: 
(1) encouraging recall by asking about the \emph{time}, \emph{location}, \emph{people}, and \emph{activity} associated with an event; 
(2) following up on keywords introduced during the conversation to deepen reminiscence; and 
(3) avoiding topics that may trigger distress. 
To operationalize these principles, \sys{} generates questions in two stages: generation and curation. 
In the generation stage, the system produces candidate questions based on the real-time conversation and context nodes in the \textit{Memory Graph}. In the curation stage, the system ranks candidates using a multi-objective scoring function: 

\[S = W_{\text{rel}} S_{\text{rel}} + W_{\text{div}} S_{\text{div}} - W_{\text{neg}} S_{\text{neg}}\]

This formulation balances conversational coherence, diversity of prompts, and avoidance of distressing content during reminiscence. 
The \textbf{relevance} term, \(S_{\text{rel}} = \mathrm{sim}(c, \mathcal{C})\), ensures that a question \(c\) is relevant to the current conversation \(\mathcal{C}\). 
The \textbf{diversity} term, \(S_{\text{div}} = 1 - \max_{k \in \mathcal{K}} \mathrm{sim}(c, k)\), prioritizes questions that differ from those in the existing question set \(\mathcal{K}\), thereby avoiding redundancy among suggestions. 
The \textbf{negative emotion} term, \(S_{\text{neg}} = \max_{t \in \mathcal{T}_{\text{neg}}} \mathrm{sim}(c, t)\), penalizes candidates that resemble known distressing topics \(\mathcal{T}_{\text{neg}}\), thereby avoiding triggering negative emotions. 
All similarity scores are computed using cosine similarity over embeddings generated by Qwen3-Text-Embedding-V4 \cite{zhang2025qwen3embedding}. 
Empirically, we set \(W_{\text{rel}} = W_{\text{\xsc{div}}} = 1\) and \(W_{\text{neg}} = 100\). 
\xsc{
We determined these weights through iterative testing with two therapists (F1 and F5 in Table~\ref{tab:formative_demographics}). 
Starting from equal weights, we adjusted them based on therapist feedback until both therapists agreed that the generated questions were of high overall quality. 
The final large distress penalty reflects therapists' preference to avoid distressing questions. }
Based on iterative testing with therapists, the system generates six candidate questions every ten seconds, selects the top three after curation, and appends them to the question list, enabling real-time access to high-quality questions.

\subsubsection{\textbf{Relevant Photos}} 
Relevant photos help therapists adapt their guidance to the PwD's real-time interests. These photos are retrieved from the \textit{Memory Graph} and selected based on their relevance to the conversational context. 
For each photo \(p\) in the \textit{Memory Graph}, we compute a relevance score: \(S_{\mathrm{photo}}(p)=\max_{n\in\mathcal{N}(p)} \mathrm{sim}(n,\mathcal{C})\), where \(\mathcal{N}(p)\) is the set of context nodes associated with \(p\) and \(\mathcal{C}\) denotes the current conversation context. 
A photo receives a high score if any of its associated context nodes is highly relevant to the conversation. 
Based on iterative testing with therapists, the top three ranked photos are presented in an auto-scrolling list that updates every 20 seconds. 
Therapists can click any photo to display it to the PwD.

\subsubsection{\textbf{Sensitivity Prompts}} 
\sys{} detects two types of sensitive situations that commonly arise in reminiscence conversations with PwD \cite{park2019systematic}: \textit{off-topic responses} and \textit{distressing topics}. 
\textbf{Off-topic responses} are detected when the conversation drifts from the session's theme. 
In such cases, the system highlights a theme-relevant guiding question in orange to help the therapist redirect the discussion. 
\textbf{Distressing topics} are detected by comparing the conversation against a predefined list of distressing events, such as divorce, bereavement, and job loss, as well as additional sensitive topics provided by the PwD's family members. When such topics arise, the system recommends switching to a different photo to redirect the conversation, in line with recommended practice \cite{RT_handbook_2005}. 
Both types of situations are detected using Qwen-Flash \cite{yang2025qwen3}, which evaluates the conversation every 20 seconds. 
\xsc{To reduce overly sensitive alerts, we evaluated thresholds of one, two, or three consecutive detections with two therapists. Both agreed that two consecutive detections provided the best balance between avoiding alerts triggered by brief mentions and minimizing alert delays. We thus adopted this threshold in the system.} 
These sensitivity prompts are shown in the same list as the guiding questions, allowing therapists to notice them quickly without shifting their visual focus.

\subsection{Session Summary Generation}
After each session, \sys{} generates structured summaries of the reminiscence conversation. 
For each photo, it produces a summary of the memories shared by the PwD along with the PwD’s emotional responses. 
Figure~\ref{fig:walkthrough}(C) illustrates an example. 
These outputs are generated by Qwen3-Max \cite{yang2025qwen3}, which is prompted to summarize the key information from the conversation and infer likely emotional responses from the sentiment expressed in the transcript. 
Therapists can further revise both the summaries and the recorded emotions to document the PwD's life story. 
These records may also support future interventions by indicating which topics the PwD is more comfortable discussing and which may be better avoided.

\subsection{Implementation Details}
The \sys{} pipeline is implemented in Python 3.12. 
Its user interface, developed using Flask, HTML, and vanilla JavaScript, runs in a web browser on a therapist-facing laptop. 
The photo currently under discussion is displayed on a separate screen for the PwD, allowing them to focus on the image without distraction. 
To minimize the risk of evoking distressing memories, the PwD's family members are asked to exclude potentially upsetting photos and provide a list of sensitive topics in advance. 
To safeguard user privacy, all cloud API requests are routed through services that adhere to a no-training, no-storage data policy. 
The prompts used in the system are provided in the supplementary materials.



\section{User Evaluation}
We conducted a within-subject field study with eight therapist-PwD dyads to assess \sys{}'s effectiveness in supporting both the \textit{planning} and \textit{facilitation} of real-world PRT sessions.

\subsection{\textbf{Participants and Recruitment}} 
To ensure the ethical appropriateness of the recruitment process, we adopted a two-stage recruitment strategy in which therapists were recruited first and then asked to help identify appropriate PwD for participation. 
After obtaining approval from our institution's IRB, we began by recruiting therapists through dementia care institutions that had long collaborated with our research team, supplemented by snowballing, \xsc{with the criteria of having at least one year of experience in conducting reminiscence therapy with PwD}. Enrolled therapists were then asked to assist in recruiting suitable PwD participants, \xsc{with the criteria of (1) being diagnosed with early-stage dementia, and (2) having the basic cognitive and communicative abilities for RT, including understanding prompts, recalling personal experiences, and engaging in conversation.}

A total of eight therapist-PwD dyads were finally recruited (Table~\ref{tab:participants_demographics}). 
Among eight therapists, one was male and seven were female, with ages ranging from 20 to 39 years (M = 25.8), and experience in dementia care ranging from 1 to 10 years (M = 3.1). 
Among the PwD, two were male and six were female, with ages ranging from 72 to 95 years (M = 84.3). 
All had been diagnosed with early-stage dementia. 
Six of them lived in care facilities, and two lived at home. 
The predominance of female participants in this sample is consistent with the higher representation of women among both PwD \cite{who_dementia_2025} and healthcare workers \cite{who2021gender}. 
None of the therapists had participated in the formative study.

\subsection{\textbf{Study Preparation}}
Before the study, we prepared the photo materials and established the comparison condition used in the evaluation. 

\paragraph{\textbf{Photo Collections}}
For each PwD, family members provided a collection of personal photos based on two criteria: 
(1) the photos represented diverse life periods, people, places, and events, and 
(2) to the best of the family members' knowledge, the photos did not include content associated with distressing memories. 
The collections ranged from 21 to 106 photos (M = 57.6). 
Family members also provided a list of negative life experiences (e.g., divorce) to help avoid sensitive topics during conversations.

\paragraph{\textbf{Baseline}}
For comparison, we implemented a baseline system reflecting therapists' current practices, as identified in our formative study and prior work \cite{woods2018reminiscence}. 
In this system, therapists manually created session plans by selecting photos and adding questions and notes for each image (see Figure~\ref{fig:baseline}). 
Both systems used the same photo collection to maintain study control. 
The key difference between them is that \sys{} provides AI-generated plans during planning and context-aware suggestions during facilitation, whereas the baseline displays therapist-authored questions.

\subsection{\textbf{Study Procedure and Data Collection}}
The study began with a 10-minute tutorial in which therapists were introduced to both systems. 
After confirming that they were comfortable using them, they proceeded to the two study phases, each designed to evaluate one of \sys{}'s two core functions. 

\paragraph{\textbf{\textit{Planning Phase.}} }
Therapists first reviewed the photo collection and identified two themes for the two PRT sessions. 
These themes were then randomly assigned to the two systems. 
Therapists used each system to prepare a session plan, with system order counterbalanced across participants to control for order effects. 
They were free to use any available features and stopped once they considered the plan ready for use. 
After completing both plans, therapists rated the plan quality, preparation efficiency, and perceived workload using the NASA-TLX \cite{hart1988development}. 
All measures used a 7-point Likert scale. 
Finally, therapists participated in a 15-minute semi-structured interview about the benefits and limitations of the two systems.

\paragraph{\textbf{\textit{Facilitation Phase.}} }
After planning, therapists conducted two PRT sessions with the PwD. 
Sessions took place in PwD's home to maximize comfort. 
System order was counterbalanced across dyads. 
Therapists were instructed to conduct the PRT sessions as they normally would and to use system features as needed. 
To ensure safety, therapists could pause or terminate a session at any time. 
Researchers observed the sessions without interfering. 
Each session typically lasted 15 to 30 minutes, with a 5-minute break between sessions. 
After completing both sessions, therapists completed the NASA-TLX \cite{hart1988development} to assess perceived task load and rated perceived conversation quality, agency, and the quality of context-aware suggestions. 
They then participated in a 15-minute semi-structured interview to reflect on their overall experiences and to discuss the benefits and limitations of both systems.

\subsection{\textbf{Data Analysis}}
We collected audio recordings, interaction logs, and questionnaire responses during the study. 
\sta{Both objective measures satisfied the Shapiro--Wilk normality assumption: planning time ($W = .99$, $p = .98$) and conversation duration ($W = .94$, $p = .67$). 
Accordingly, these measures were analyzed using paired $t$-tests, with Cohen's $d_z$ reported as the effect size. 
Subjective ratings were analyzed using Wilcoxon signed-rank tests \cite{woolson2007wilcoxon}, with $r$ reported as the effect size.} 
Interview recordings were transcribed and analyzed using thematic analysis \cite{clarke2017thematic}. 
\xsc{
Two authors independently coded transcribed data, compared initial codes, and resolved discrepancies through iterative discussion. They then grouped these codes into higher-level themes and reviewed these themes to ensure that they accurately captured the key findings.} 
\sta{Given the small sample size and multiple comparisons, we treat the statistical findings as indicative trends and interpret them alongside the qualitative findings.} 


\subsection{\textbf{Ethical Considerations}}

Given the particular vulnerability of PwD, we took careful steps throughout the study to protect participants' rights and privacy. 
Before each study activity, we carefully explained the study objectives as well as the procedures for collecting, storing, and using personal data to therapists and PwD's family members, obtaining their informed consent. 
During the study, therapists remained responsible for assessing PwD's condition and could pause or terminate a session at any time if signs of fatigue, discomfort, or distress emerged. 
Researchers observed the sessions without interfering unless intervention was necessary for safety. 
We also took precautions to avoid potentially distressing topics by asking family members in advance to identify sensitive life experiences and photos that should not be used. All collected data were anonymized prior to analysis, and identifying information was removed or replaced.

\section{User Evaluation Results}
In the planning phase, the eight therapists developed a total of 16 session plans. 
In the facilitation phase, one dyad (T3-P3) withdrew early from the study due to P3's physical health condition. 
The other seven dyads completed a total of 14 PRT sessions. 
Below, we report the evaluation results of \sys{} in these two phases. 

\begin{figure*}[!h]
    \centering
    \includegraphics[width=1.0\linewidth]{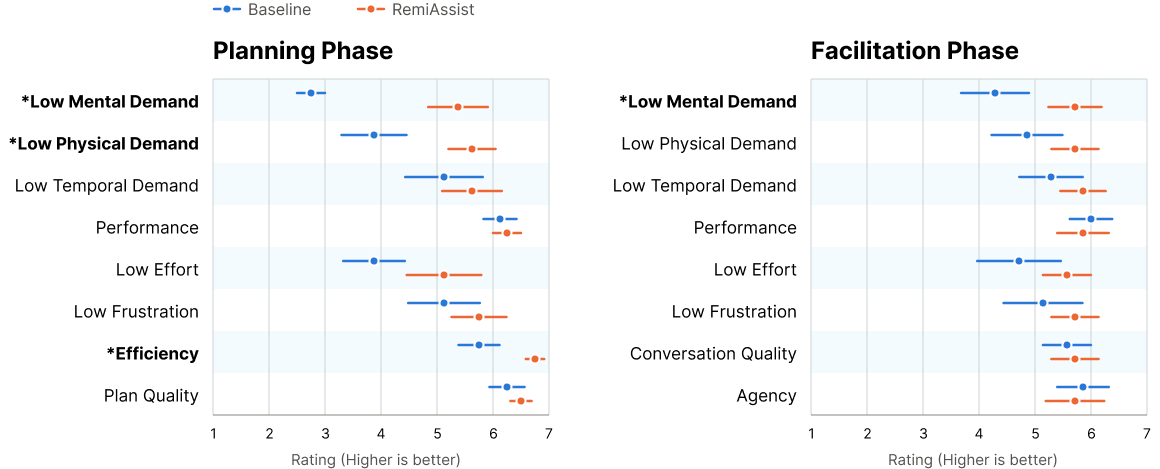}
    \caption{\xsc{Therapists' ratings for both systems (1 = strongly disagree, 7 = strongly agree)}. \xsc{Error bars indicate standard errors.} Asterisks indicate statistical significances based on the Wilcoxon signed-rank test (* denotes $\bm{p<.05}$).}
    \label{fig:rating}
\end{figure*}

\begin{figure*}[!h]
    \centering
    \includegraphics[width=0.88\linewidth]{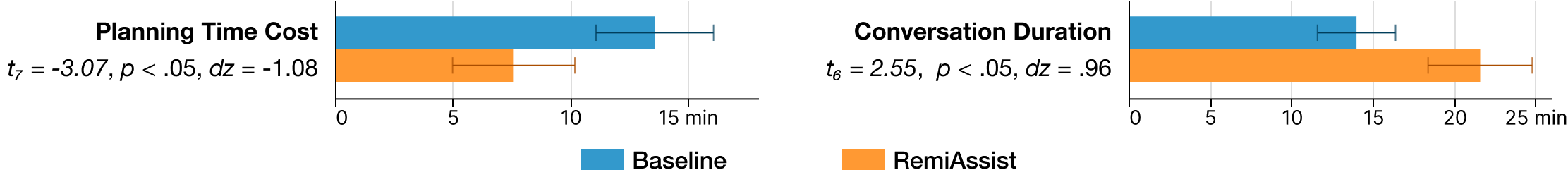}
    \caption{\sys{} reduced therapists' session planning time and helped them sustain reminiscence conversations compared with the baseline. 
    Statistical significance was assessed using paired t-tests. Error bars indicate standard errors.}
    \label{fig:evaluation_time}
\end{figure*}

\subsection{Session Planning}
\sta{\sys{} was associated with improved planning efficiency and lower perceived workload.} Detailed results are presented below.

\paragraph{\textbf{\sta{Planning Efficiency.}}} 
\sta{On average, therapists completed a session plan in 7.6 minutes with \sys{}, a 44\% reduction from the baseline average of 13.5 minutes ($t_7 = -3.07,\ p < .05$, $d_z=-1.08$)}. 
\sta{Therapists' ratings aligned with this trend, indicating higher perceived efficiency with \sys{} ($Z = -2.33,\ p < .05$, $r=0.83$)}. 
The number of photos included in the session plans did not differ significantly between conditions (\sys{}: $\mu=9.4$ v.s. baseline: $\mu=9.1$, $t_7 = 0.23,\ p = .82$, $d_z=0.08$). 
In addition, session plans created with \sys{} were rated as highly theme-focused ($\mu = 6.50$ on a 7-point scale), rich in contextual detail ($\mu = 6.38$), and high in overall quality ($\mu = 6.50$). 
Together, \sta{these results suggest} that \sys{} may improve efficiency without reducing plan quality.

\paragraph{\textbf{\sta{Perceived Planning Workload.}}} 
\sta{When using \sys{}, therapists reported lower mental demand ($Z = -2.53,\ p < .05$, $r=0.89$) and physical demand ($Z = -2.41,\ p < .05$, $r=0.85$) compared to the baseline.} 
As T5 noted, ``\textit{There is already a draft for me, so it takes much less effort to select photos and prepare questions.}'' 
Moreover, AI-inferred details helped therapists notice key information they might otherwise have overlooked, indirectly reducing their planning workload. 
As T1 explained, ``\textit{I didn't realize the photo was from a birthday because the candle was easy to miss, but the AI-inferred detail helped me notice it. It helped me understand the moment and prepare guiding questions more easily.}'' 
We also evaluated the quality of the AI-generated plans, as reported in Section~\ref{sec:tech_eval}.

\subsection{Real-Time Facilitation}
\sta{\sys{} was associated with longer reminiscence conversations and lower cognitive load.} Therapists also reported several benefits of the context-aware suggestions. We describe these findings below.


\paragraph{\textbf{\sta{Conversation Duration.}}} 
\sta{\sys{}-mediated reminiscence sessions lasted an average of 21.6 minutes, representing a 54\% increase over the 14.0-minute baseline ($t_6 = 2.55,\ p < .05$, $d_z=0.96$).} 
Our participants attributed this increase to \sys{}'s ability to sustain conversational flow and deepen engagement. As T7 noted, ``\textit{Its relevant suggestions helped me follow up on the PwD's interests and deepen the conversation}''. 
Therapists did not report a significant difference in perceived conversation quality between the two conditions ($Z=-1.00,\ p=0.32$, \sta{$r=0.38$}). 
They also reported no significant reduction in agency compared to the baseline ($Z=-0.48,\ p=0.66$, \sta{$r=0.17$}), describing \sys{}'s suggestions as ``\textit{suggestive rather than directive}'' (T2). 
During the conversations, all therapists used \sys{} on demand rather than monitoring it continuously. 
This enabled them to prioritize natural communication with the PwD and consult the system only when needed. 
As T8 noted, ``\textit{Natural communication with the PwD is always the first priority, so I mainly focused on the PwD and glanced at the system on demand.}''

\paragraph{\textbf{\sta{Perceived Cognitive Load.}}} 
\sta{Compared to the baseline, \sys{} was associated with therapists' lower mental demand during facilitation ($Z=-2.23,\ p<.05$, $r=0.84$)}. 
Therapists primarily attributed this benefit to the system's always-available support and its ability to help them manage uncertainty during sessions. 
As T5 explained, ``\textit{There are many unpredictable situations in RT: a person may have difficulty focusing on the conversation or may become distressed. 
Having suggestions readily available helps us manage these situations.}'' 
Similarly, T2 noted, ``\textit{Sometimes the PwD goes far off topic, and I'm not always sure how to guide the conversation back. The system's suggestions help me get it back on track more easily.}''

\paragraph{\textbf{Benefits of Context-Aware Suggestions.}}
Therapists reported that the three types of suggestions offered distinct benefits during different session stages. 
\textbf{Guiding questions} helped therapists initiate conversations and sustain deeper discussion over time. As T5 explained, ``\textit{When switching to a new photo, the guiding questions helped me smoothly start the conversation. As the discussion went on, I also referred back to the questions, compared them with my initial thoughts, and identified more meaningful topics for discussion.}'' 
\textbf{Relevant photos} supported therapists in adapting to the PwD's changing interests in real time, which helped maintain engagement. For example, T6 noted, ``\textit{Initially, I mainly prepared photos about family members, but I felt that she was more interested in her schoolmates, so I showed those photos to encourage her to share more.}'' 
\textbf{Sensitivity prompts} provided therapists with timely support for managing vulnerable moments by enabling smooth topic redirection. As T2 described, ``\textit{When distressing topics came up, I could easily panic. Clicking on the prompt to switch to another photo was the easiest way to redirect the PwD's attention.}'' 
Additionally, we evaluated the quality of context-aware suggestions, as reported in Section~\ref{sec:tech_eval}.

\section{Technical Evaluation}\label{sec:tech_eval}
We evaluated \sys{}'s technical performance in two areas: (1) the quality of generated session plans and (2) the quality of context-aware suggestions. 

\paragraph{\textbf{Session Plan Quality.}} 
\xsc{We evaluated the accuracy and usefulness of the generated session plans via independent expert review. 
To assess accuracy, one therapist annotated errors in all unedited plans, and a second reviewed the annotations for reliability. Accuracy rates were 91\% for photo selection, 90\% for guiding questions, and 93\% for contextual details, indicating consistently high accuracy across content types. 
To assess usefulness, two therapists, blinded to editing conditions, independently rated pre- and post-edit plans on a 7-point scale. Mean ratings increased from 5.8 to 6.4 after editing, indicating that the generated plans were fairly useful before editing.} 
We further analyzed how therapists revised the generated plans. 
Across eight revised plans, therapists added or removed 12 of 75 photos (16.0\%) and revised 22 of 150 questions (14.7\%). 
Interview data suggest that these edits mainly revised inaccuracies caused by the system's limited access to personal context. 
For example, T1 noted, ``\textit{We will talk about families, but I know some photos depict her colleagues rather than family members, so I removed those photos.}'' 
Therapists also revised guiding questions to avoid unverified assumptions. As T5 explained, ``\textit{The question `Was it winter?' assumes this photo was taken in winter, but that is not clear. It is safer to simply ask which season it was.}'' 
Beyond correcting inaccuracies, therapists also reordered photos to create more coherent narratives. 
As T3 explained, ``\textit{For a love story, I might arrange the photos chronologically, whereas for people, I might group them by relationship.}'' 
These findings suggest that session planning depends on therapists’ personal knowledge of the PwD and their ability to organize materials into meaningful narratives. 
We discuss design implications for supporting this process in Section~\ref{sec:long-term_profile}.

\paragraph{\textbf{Context-Aware Suggestion Quality}} 
We evaluated context-aware suggestion quality by asking therapists to rate how well each type of suggestion supported their real-time facilitation needs. 
Ratings were high for both guiding questions ($\mu = 6.29, \ \sigma=0.76$ on a 7-point scale) and relevant photos ($\mu = 6.14, \ \sigma=0.90$), whereas sensitivity prompts received a slightly lower rating ($\mu = 5.86, \ \sigma=1.35$). 
Interview data suggest that this lower rating mainly reflected differences in therapists' professional judgment about when intervention is appropriate. 
As T8 noted, ``\textit{Sometimes off-topic or sad memories are not necessarily bad. The PwD may want to share their feelings in the moment. So I felt the notification was more cautious than I expected.}'' 
This suggests that sensitivity prompts should better align with therapists' professional judgment, which we discuss in Section~\ref{sec:safe_ethical_discussion}. 
Therapists also identified two key areas for improving the suggestions. 
First, they recommended expanding guiding questions beyond factual prompts to better elicit personal storytelling. As T5 explained, ``\textit{Current questions are primarily factual, such as `Where is your wedding?' But we also ask many questions that elicit personal stories, such as `Was it a traditional-style or modern-style wedding?'}'' 
Second, they noted that the AI should better interpret personal referents, such as ``\textit{my son}'' or ``\textit{my workplace}'' (T4), to retrieve relevant photos. 
This points to the value of developing long-term memory profiles, which we discuss in Section~\ref{sec:long-term_profile}. 

Additionally, we evaluated the distinguishability of themes in the \textit{Memory Graph} and the frequency of context-aware suggestions; detailed results are reported in Appendix~\ref{sec:tech_statistics}.



\section{Discussion}
Therapists are central to the effective delivery of RT for PwD, yet practical RT requires substantial time, effort, and expertise. 
\sys{} provides support to therapists through two core techniques: a \textit{Memory Graph} that enables efficient intervention planning, and a \textit{Context-Aware Guiding Strategy} that supports real-time facilitation. 
\sta{Our field study provides preliminary evidence that \sys{} was associated with lower planning workload and provided useful context-aware suggestions during real-time facilitation.} 
In the following sections, we discuss key lessons from our field study, including: 
(1) \sys{}'s effects on PwD engagement, 
(2) supporting flexible agency for therapists, 
(3) accommodating diverse communication styles, 
(4) developing long-term memory profiles, and 
(5) ensuring safe and ethical intervention.

\subsection{\xsc{\sys{}'s Effect on PwD Engagement}}\label{sec:discussion_PwD_engagement}
\xsc{
Due to ethical considerations \cite{NPI_chi24}, we did not collect self-reported data directly from PwD. Instead, we assessed \sys{}'s influence on PwD engagement through therapists' observations. 
During \sys{}-mediated sessions, six therapists observed that PwD engaged in more verbal sharing. They attributed this increase to \sys{}'s ability to sustain conversation: ``\textit{The suggestions helped keep the conversation going, so they could continue talking, which is important for maintaining cognitive function}'' (T8). 
Four therapists also reported that PwD recalled memories more coherently when the system provided cues relevant to the ongoing conversation. 
As T5 explained, ``\textit{The memory cues aligned with their interest, so they can recall more coherently}.'' 
These observations suggest that \sys{} may enhance PwD engagement during RT, although further validation is needed. 
Future work should investigate ethical approaches to assessing PwD engagement, such as measuring utterance frequency and emotional responses \cite{NPI_chi24}.
}

\subsection{Supporting Flexible Agency for Therapists}\label{sec:discussion_agency}
To reduce therapists' cognitive load during \sys{}-mediated RT, we chose not to expose the raw \textit{Memory Graph}, since interpreting graph structures could be cognitively demanding \cite{huang2009measuring}. 
Instead, the system presents curated, context-aware suggestions to the therapists. 
Yet, during the field study, some therapists desired more flexible access to the PwD's life history. 
For example, T4 wanted to understand how different themes overlapped within the PwD's photo collection, 
whereas T8 wanted to inspect relationships among photos. 
These findings indicate an opportunity for future systems to support flexible exploration of the \textit{Memory Graph} for therapists. 
One possible direction is to offer multiple representations of the \textit{Memory Graph}, such as life-event timelines \cite{aigner2023timeline_visualization}, family-tree views \cite{fu2017ancestral}, or graph visualizations \cite{zhang2025neurosync}, so that therapists can select the representation that best fits their need. 
A key open challenge, however, is how to enable such flexibility without increasing cognitive load. This may require context-aware visualizations that highlight the most relevant information based on the current context.

\subsection{Supporting Diverse Communication Styles}\label{sec:discussion_communication_style}
During practical reminiscence sessions, PwD exhibited differences in both \textit{conversational initiative} \cite{chu1997tracking} and \textit{attentional focus} \cite{parasuraman1992visuospatial}. 
Some PwD (e.g., P4) proactively shared related memories, while others (e.g., P6) mainly responded to questions. 
Likewise, some PwD are focused on the session theme, while others are more likely to drift off topic. Based on these patterns, we identify three key communication styles that future systems could support.

\textbf{When PwD are proactive and focused}, they may need less therapist guidance and benefit more from tools that preserve their agency. 
In these cases, future systems could enable them to select relevant photos during conversations \cite{whittaker2010easy,levy2023chatting}. 
These designs should also account for the PwD's digital literacy \cite{tsai2017social}. PwD with higher digital literacy may prefer digital interfaces, while those with lower digital literacy may engage more effectively with printed photos \cite{seah2026rememo}. In printed-photo settings, \sys{} could be integrated into smart glasses \cite{vid2coach_uist25,wang2026wearable} to access the current photo and provide therapists with relevant suggestions.

\textbf{When PwD are proactive but easily distracted}, they may participate actively but be more likely to drift off topic. 
In these cases, systems could help therapists redirect attention and sustain engagement by providing multimodal stimuli, such as relevant photos, music \cite{music_RT_chi24}, and videos \cite{xu2025branch,xu2026sonic}.

\textbf{When PwD are reactive and share few memories}, 
they tend to rely more heavily on therapists to sustain the conversation. 
In these cases, future systems could provide stronger conversational scaffolding, such as through memory-elicitation strategies that encourage recall. 
Importantly, these communication styles should not be treated as fixed individual traits; they are shaped by situated context (e.g., PwD's current mood, health condition, etc.). Future systems should therefore be designed to detect shifts in communication styles and adapt their support accordingly.

\subsection{Developing Long-Term Memory Profiles}\label{sec:long-term_profile}
In \sys{}, we introduced the \textit{Memory Graph} to represent a PwD's personal memory profile. 
In future systems, this profile could be continuously updated across reminiscence sessions to support longitudinal care. 
As suggested by therapists, such a profile could record three key types of information. 
First, it should preserve the PwD's \textbf{contextual history}, such as ``\textit{this is my son}'' or ``\textit{that is my hometown}'', by annotating photos during sessions (T8). 
Second, it should model the PwD's \textbf{emotional responses} to different topics. AI may help infer these patterns from conversations, with therapists validating the results to ensure safe and ethical use. 
Third, it should record effective \textbf{comfort strategies}, such as ``\textit{which photos or topics help quickly soothe the PwD during distress}'' (T5). 
Beyond therapists, family members could also annotate photos and verify personal context, helping ensure the information reliability.

Furthermore, future systems must address both technical and privacy challenges. From a technical perspective, they need mechanisms to update the \textit{Memory Graph} when new information conflicts with prior knowledge \cite{zhong2024memorybank}, and to adapt future session plans and context-aware guidance based on earlier conversations. From a privacy perspective, these profiles require strong safeguards. Future work should explore privacy-preserving approaches, such as local AI models \cite{chu2024mobilevlm}, to better protect sensitive personal data.

\subsection{Ensuring Safe and Ethical Intervention}\label{sec:safe_ethical_discussion}
Reminiscence is ethically sensitive because it can evoke both positive and painful emotions \cite{bittersweet_chi25}. 
In real-world interventions, ensuring ethical safety is therefore critical, especially when painful emotions are triggered. 
We observed that therapists responded to these moments differently: 
T8 chose to ``\textit{let the PwD share his feelings}'', while 
T5 first offered comfort and then changed the photo to redirect the conversation. 
The differences reflect therapists' professional judgment. 
As T8 explained, ``\textit{I felt he wanted to share more, so it was better to listen attentively and acknowledge his feelings.}'' 
These responses highlight that therapists' professional judgment is central to safe and ethical intervention. 

Based on therapists' feedback, we identify three ways in which AI could support safe interventions. 
First, AI could summarize previously discussed topics and related emotions to help therapists identify appropriate topics and avoid unsuitable ones. 
Second, during distressing moments, AI could help therapists quickly retrieve comforting photos. 
\xsc{Third, AI could adapt sensitivity-prompt detection thresholds to individual therapists' judgment styles by learning from their responses over time \cite{wang2024comprehensive}. Such adaptation could better align system behavior with therapists' professional judgments.} 

\subsection{Limitations and Future Work}
We discuss limitations and future work in two areas: (1) the scope of \sys{} and (2) the design of the user study.

\paragraph{\textbf{Scope of \sys{}.}} 
\xsc{\sys{} currently focuses on people with early-stage dementia, who retain the basic cognitive and communicative abilities needed to engage in RT \cite{cuevas2020reminiscence}. Future work should examine how \sys{} can accommodate changing abilities across dementia stages, including reduced verbal fluency in the middle and late stages \cite{kindell2017everyday}.} 
While \sys{} focuses on photo-based RT, future work could support other reminiscence materials, such as music, video, or virtual reality \cite{music_RT_chi24,xu2025danmua11y,relive_vr2025}. 
Beyond supporting therapists, future work could also examine how \sys{} might support family members in facilitating RT. 
Finally, the AI-inferred contextual details in \sys{} are intended only as therapist-facing references and should not be presented directly to PwD, as inaccuracies could distort their memories \cite{pataranutaporn2025synthetic}.

\paragraph{\textbf{Limitations of the User Study.}} 
\xsc{Our field study has several limitations. 
First, the small sample size may have limited statistical power, while uncorrected multiple comparisons may have inflated the risk of Type I error \cite{rifat2024cohabitant}. Thus, the statistical significance should be interpreted as exploratory rather than confirmatory. 
Second, the short-term, within-subject design may have introduced novelty and carryover effects. 
Larger-scale, longitudinal studies are needed to assess the long-term effects of the system. 
Finally, while comparing \sys{} with therapists' current practices enabled us to assess its practical utility, the comparison did not isolate context awareness from differences in information quantity. 
Future studies should include an equal-information, non-context-aware baseline to isolate the contribution of the context-aware design.
}



\section{Conclusion}
We have presented \sys{}, a system that supports therapist-in-the-loop photo-based reminiscence therapy in dementia care. 
It integrates a \textit{Memory Graph} for theme-centered intervention planning and a \textit{Context-Aware Guiding Strategy} for real-time session facilitation. 
\sta{Findings from our field study indicate that the system was associated with lower workload among therapists in both phases.} 
Our study further highlights broader opportunities for context-aware systems to augment therapist-PwD communication, including supporting diverse communication styles, developing long-term memory profiles, as well as designing safe and ethical interventions. 
We hope this work provides insights into the design of AI systems that empower therapists and inspires future work on more personalized reminiscence therapy in dementia care.


\begin{acks}
The authors thank all participants for their support during the studies and are grateful to the reviewers for their constructive feedback. 
Shuchang Xu also expresses his gratitude to his grandmother, Shuxiang Sun, for inspiring this work.
\end{acks}

\bibliographystyle{ACM-Reference-Format}
\bibliography{references}

\appendix

\newpage
\onecolumn


\section{Implementation Details of the Baseline System}\label{sec:baseline_details}
\begin{figure*}[!h]
    \centering
    \includegraphics[width=\linewidth]{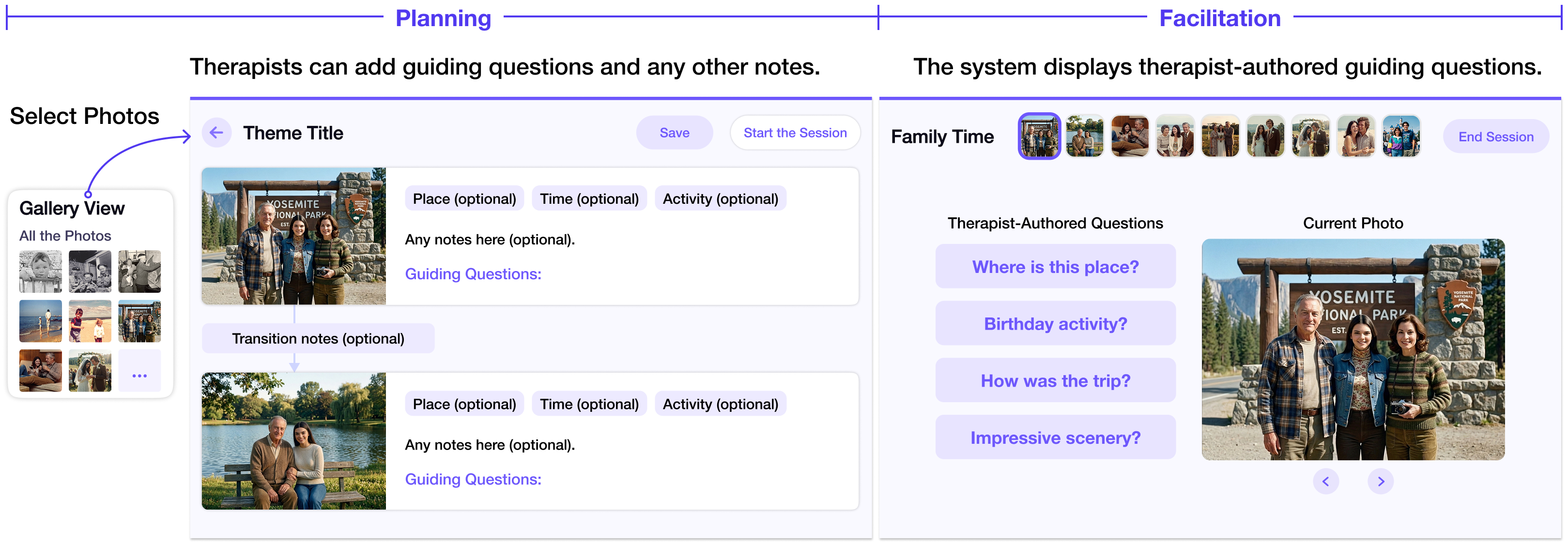}
    \caption{The baseline system is designed to reflect therapists' manual PRT practice. During the preparation phase, therapists select photos from a gallery view and annotate each image with guiding questions and notes. During the facilitation phase, the system displays the guiding questions authored by the therapists.}
    \label{fig:baseline}
\end{figure*}

The baseline system is designed to reflect therapists' manual PRT practice \cite{woods2018reminiscence}. 
Figure~\ref{fig:baseline} shows the user interfaces of the baseline system. 
During the planning phase, therapists select photos from a gallery view and annotate each image with guiding questions and notes. 
During the facilitation phase, the system displays the guiding questions authored by the therapists. 

\section{Technical Performance of the \sys{} Pipeline}\label{sec:tech_statistics}
Based on data from the evaluation study, we report technical statistics for the \sys{} pipeline in two areas: (1) theme distinguishability in the \textit{Memory Graph}, and (2) the frequency of context-aware suggestions.

\paragraph{\textbf{Theme Distinguishability in the \textit{Memory Graph}}} 
On average, each \textit{Memory Graph} contained 57.6 photos (SD = 32.3) and 5.5 themes (SD = 2.6). To assess theme distinguishability, we compared edge weights for photo pairs within the same theme versus across different themes. The mean within-theme edge weight was 2.80 ($\sigma = 0.80$), which was 1.61 times higher than the mean between-theme edge weight of 1.74 ($\sigma = 0.24$). 
These results suggest that \sys{} effectively groups photos into \textit{internally coherent}, \textit{externally distinct} themes.

\paragraph{\textbf{Frequency of Context-Aware Suggestions}}
Table~\ref{tab:context_aware_suggestions} shows the average number of context-aware suggestions generated per session. Guiding questions were the most common type of context-aware suggestion, followed by relevant photo suggestions, whereas sensitivity prompts occurred least often. 
Sensitivity prompts were relatively rare because they were triggered primarily when PwD shifted the conversation substantially away from the reminiscence theme (e.g., toward religious preaching), as well as in emotionally sensitive contexts (e.g., bereavement, divorce, difficult life periods, and negative interpersonal relationships).

\aptLtoX{\begin{table}[h]
\centering
\caption{Theme distinguishability statistics.}
\label{tab:theme-distinguishability}
\begin{tabular}{c  c  c }
\toprule
\textbf{Statistic} & \textbf{Mean} & \textbf{SD} \\
\midrule
Photos per collection & 57.6 & 32.3 \\
Themes per collection & 5.5 & 2.6 \\
\midrule
Within-theme edge weight & 2.80 & 0.80 \\
Between-theme edge weight & 1.74 & 0.24 \\
\bottomrule
\end{tabular}
\end{table}

\begin{table}
\centering
\caption{Number of context-aware suggestions per session.}
\label{tab:context_aware_suggestions}
\renewcommand{\arraystretch}{1.3}
\begin{tabular}{lcc}
\toprule
\textbf{Suggestion Type} & \textbf{Mean} & \textbf{SD} \\
\midrule
Guiding questions & 167.3 & 43.2 \\
Relevant photos & 35.7 & 14.4 \\
Sensitivity prompts & 7.7 & 5.0 \\
\bottomrule
\end{tabular}
\end{table}}{\begin{table}[h]
\centering

\begin{minipage}[t]{0.48\textwidth}
\centering
\caption{Theme distinguishability statistics.}
\label{tab:theme-distinguishability}
\begin{tabular}{c  c  c }
\toprule
\textbf{Statistic} & \textbf{Mean} & \textbf{SD} \\
\midrule
Photos per collection & 57.6 & 32.3 \\
Themes per collection & 5.5 & 2.6 \\
\midrule
Within-theme edge weight & 2.80 & 0.80 \\
Between-theme edge weight & 1.74 & 0.24 \\
\bottomrule
\end{tabular}
\end{minipage}
\hfill
\begin{minipage}[t]{0.5\textwidth}
\centering
\caption{Number of context-aware suggestions per session.}
\label{tab:context_aware_suggestions}
\renewcommand{\arraystretch}{1.3}
\begin{tabular}{lcc}
\toprule
\textbf{Suggestion Type} & \textbf{Mean} & \textbf{SD} \\
\midrule
Guiding questions & 167.3 & 43.2 \\
Relevant photos & 35.7 & 14.4 \\
Sensitivity prompts & 7.7 & 5.0 \\
\bottomrule
\end{tabular}
\end{minipage}
\end{table}}

\newpage
\section{Rating Statistics in the Evaluation Study}


\begin{table*}[!hbtp]
\caption{\centering Statistics of subjective ratings in the evaluation study. \xsc{Scale direction: 1 = strongly disagree; 7 = strongly agree}.\newline 
SD denotes the standard deviation. Significance was analyzed using the Wilcoxon signed-rank test.}
\label{tab:subjective_ratings}
\centering
\resizebox{1.0\columnwidth}{!}{
\setlength{\tabcolsep}{1.15mm}{
\renewcommand\arraystretch{1.4}
\newcommand{\hlineblack}{\specialrule{0.1em}{0em}{0em}}
\begin{tabular}{c | cc | cc | c | c | cc | cc | c}
\hlineblack
\multicolumn{6}{c|}{\textbf{Preparation Phase}} & \multicolumn{6}{c}{\textbf{Facilitation Phase}} \\
\cline{1-12}
\multirow{2}{*}{\textbf{Aspect}} & \multicolumn{2}{c|}{\textbf{Baseline}} & \multicolumn{2}{c|}{\textbf{\sys{}}} & \multirow{2}{*}{\textbf{Significance}}
& \multirow{2}{*}{\textbf{Aspect}} & \multicolumn{2}{c|}{\textbf{Baseline}} & \multicolumn{2}{c|}{\textbf{\sys{}}} & \multirow{2}{*}{\textbf{Significance}} \\
\cline{2-5} \cline{8-11}
& \textbf{Mean} & \textbf{SD} & \textbf{Mean} & \textbf{SD} &
& & \textbf{Mean} & \textbf{SD} & \textbf{Mean} & \textbf{SD} & \\
\hlineblack

\xsc{Low} Mental Demand   & 2.75 & 0.71 & 5.37 & 1.51 & $Z=-2.53,\ p<.05$
& \xsc{Low} Mental Demand        & 4.29 & 1.60 & 5.71 & 1.25 & $Z=-2.23,\ p<.05$ \\

\xsc{Low} Physical Demand    & 3.87 & 1.64 & 5.62 & 1.19 & $Z=-2.41,\ p<.05$
& \xsc{Low} Physical Demand     & 4.86 & 1.68 & 5.71 & 1.11 & $Z=-1.89,\ p=.06$ \\

\xsc{Low} Temporal Demand & 5.12 & 1.96 & 5.62 & 1.51 & $Z=-1.41,\ p=.16$
& \xsc{Low} Temporal Demand      & 5.29 & 1.50 & 5.86 & 1.07 & $Z=-1.63,\ p=.10$ \\

Performance     & 6.13 & 0.84 & 6.25 & 0.71 & $Z=-1.00,\ p=.32$
& Performance          & 6.00 & 1.00 & 5.86 & 1.22 & $Z=-1.00,\ p=.32$ \\

\xsc{Low} Effort         & 3.87 & 1.55 & 5.12 & 1.89 & $Z=-1.62,\ p=.11$
& \xsc{Low} Effort               & 4.71 & 1.98 & 5.57 & 1.13 & $Z=-1.60,\ p=.11$ \\

\xsc{Low} Frustration     & 5.12 & 1.81 & 5.75 & 1.39 & $Z=-1.51,\ p=.13$
& \xsc{Low} Frustration          & 5.14 & 1.86 & 5.71 & 1.11 & $Z=-1.13,\ p=.26$ \\

Efficiency      & 5.75 & 1.04 & 6.75 & 0.46 & $Z=-2.33,\ p<.05$
& Conversation Quality & 5.57 & 1.13 & 5.71 & 1.11 & $Z=-1.00,\ p=.32$ \\

Plan Quality    & 6.25 & 0.89 & 6.50 & 0.54 & $Z=-1.41,\ p=.16$
& Agency               & 5.86 & 1.22 & 5.71 & 1.38 & $Z=-0.45,\ p=.66$ \\

\hlineblack
\end{tabular}
}
}
\end{table*}


\section{Participant Demographics}


\begin{table}[h]
\centering
\caption{Participant demographics in the formative study. F1--F5 refer to therapists.}
\label{tab:formative_demographics}
\renewcommand\arraystretch{1.4}
\setlength\tabcolsep{10pt}
\newcommand{\hlineblack}{\specialrule{0.1em}{0em}{0em}}
\begin{tabular}{cccc}
\hlineblack
\textbf{ID} & \textbf{Age} & \textbf{Gender} & \textbf{Dementia Care Experience} \\
\midrule
F1 & 29 & Female & 4 Years \\
F2 & 58 & Male & 12 Years \\
F3 & 42 & Female & 8 Years \\
F4 & 22 & Male & 1 Year \\
F5 & 28 & Female & 2 Years \\
\hlineblack
\end{tabular}
\end{table}


\begin{table*}[!h]
    \centering
    \caption{\centering Participant demographics in the evaluation study. T1--T8 refer to therapists, and P1--P8 refer to people with dementia. 
    \newline Matching numbers indicate paired participants.}
    \label{tab:participants_demographics}

\renewcommand{\arraystretch}{1.4}
\renewcommand\tabularxcolumn[1]{m{#1}}

\begin{tabularx}{0.9\linewidth}{l c
    >{\hsize=0.65\hsize\centering\arraybackslash}X
    >{\hsize=1.0\hsize\centering\arraybackslash}X |
    c c
    >{\hsize=0.65\hsize\centering\arraybackslash}X
    >{\hsize=1.7\hsize\centering\arraybackslash}X
    >{\hsize=1.0\hsize\centering\arraybackslash}X}

    \hline
    \multicolumn{4}{c|}{\textbf{Therapists}} & \multicolumn{5}{c}{\textbf{People with Dementia (PwD)}} \\
    \cline{1-4} \cline{5-9}
    \textbf{ID} & \textbf{Age} & \textbf{Gender} & \textbf{Dementia Care Experience} &
    \textbf{ID} & \textbf{Age} & \textbf{Gender} & \textbf{Cognitive Condition} & \textbf{Place of Residence} \\
    \hline

    T1 & 26 & Female & 2 Years  & P1 & 72 & Female & Early-Stage Dementia & Home \\
    T2 & 22 & Female & 1 Year   & P2 & 94 & Female & Early-Stage Dementia & Care Facility \\
    T3 & 22 & Female & 1 Year   & P3 & 86 & Female & Early-Stage Dementia & Care Facility \\
    T4 & 26 & Male   & 2 Years  & P4 & 72 & Male   & Early-Stage Dementia & Home \\
    T5 & 30 & Female & 7 Years  & P5 & 86 & Female & Early-Stage Dementia & Care Facility \\
    T6 & 20 & Female & 1 Year   & P6 & 88 & Female & Early-Stage Dementia & Care Facility \\
    T7 & 21 & Female & 1 Year   & P7 & 95 & Female & Early-Stage Dementia & Care Facility \\
    T8 & 39 & Female & 10 Years & P8 & 81 & Male   & Early-Stage Dementia & Care Facility \\

    \hline
\end{tabularx}

\end{table*}

\newpage

\end{document}